# Node Failure Time and Coverage Loss Time Analysis for Maximum Stability vs Minimum Distance Spanning Tree based Data Gathering in Mobile Sensor Networks

[1]Natarajan Meghanathan and [2]Philip D. Mumford
[1]Jackson State University, 1400 Lynch St, Jackson, MS, USA
[2]Air Force Research Lab, Sensors Directorate, Wright Patterson AFB, OH, USA
[1]Corresponding Author E-mail: natarajan.meghanathan@jsums.edu

## ABSTRACT

*A mobile sensor network is a wireless network of sensor nodes that move arbitrarily. In this paper, we explore the use of a maximum stability spanning tree-based data gathering (Max.Stability-DG) algorithm and a minimum-distance spanning tree-based data gathering (MST-DG) algorithm for mobile sensor networks. We analyze the impact of these two algorithms on the node failure times and the resulting coverage loss due to node failures. Both the Max.Stability-DG and MST-DG algorithms are based on a greedy strategy of determining a data gathering tree when one is needed and using that tree as long as it exists. The Max.Stability-DG algorithm assumes the availability of the complete knowledge of future topology changes and determines a data gathering tree whose corresponding spanning tree would exist for the longest time since the current time instant; whereas, the MST-DG algorithm determines a data gathering tree whose corresponding spanning tree is the minimum distance tree at the current time instant. We observe the Max.Stability-DG trees to incur a longer network lifetime (time of disconnection of the network of live sensor nodes due to node failures), a larger coverage loss time for a particular fraction of loss of coverage as well as a lower fraction of coverage loss at any time. The tradeoff is that the Max.Stability-DG trees incur a lower node lifetime (the time of first node failure) due to repeated use of a data gathering tree for a longer time.*

## 1. INTRODUCTION

A mobile sensor network is a dynamically changing wireless distributed system of arbitrarily moving sensor nodes that operate under limited battery charge, memory and processing capacity. In addition, the bandwidth of these networks is also limited as well as the transmission range of the nodes is restricted to conserve the battery charge and to reduce collisions. With all of the above operating constraints, it is not a practically feasible solution to expect each of these sensor nodes to individually transmit their data (directly or through multi-hop route) to the control center, commonly called sink, which is typically located far away from the network field being monitored. In this context, several data gathering algorithms that focus on aggregating data from the individual sensor nodes through the use of a communication topology (like chain [1], cluster [2], tree [3], connected dominating set [4], and etc) have been proposed. However, almost all of these algorithms have been proposed for static sensor networks in which the sensor nodes have been assumed to remain fixed at a particular location for the entire lifetime.

The common objective of many of the data gathering algorithms for the static sensor networks has been to conserve energy and maximize the node lifetime and network lifetime. In this context, in a recent research [5], we evaluated the performance of the data gathering algorithms based on different communication topologies and observed the minimum distance-spanning tree based data gathering (MST-DG) trees to be the most energy-efficient. However, with mobility, the network topology changes dynamically with time and thus, there is a need to determine stable data gathering trees that do not break frequently. To the best of our knowledge, we have not come across any algorithm (centralized or distributed) for stable data gathering in mobile sensor networks.

In the first half of the paper, we propose a benchmarking algorithm for maximum stability data gathering (Max.Stability-DG) in mobile sensor networks such that the number of tree discoveries is the global minimum. Given the complete knowledge of the future topology changes, the Max.Stability-DG algorithm operates based on the following greedy principle: Whenever a data gathering tree is required at time instant $t$, choose the longest-living data gathering tree from $t$. The above strategy is repeated over the duration of the data gathering session. The sequence of such longest-living data gathering trees incurs the minimum number of tree discoveries. The worst-case run-time complexity of the Max.Stability-DG tree algorithm is $O(n^2 T \log n)$ and $O(n^3 T \log n)$ when operated under sufficient-energy and energy-constrained scenarios respectively, where $n$ is the number of nodes in the network and $T$ is the total number of rounds of data gathering; $O(n^2 \log n)$ is the worst-case run-time complexity of the minimum-weight spanning tree algorithm (we use Prim's algorithm [6]) used to determine the underlying spanning trees from which the data gathering trees are derived. A similar approach is adopted to determine the sequence of MST-DG trees – with the only difference being that the underlying spanning tree is a minimum distance spanning tree determined based on the local network topology and not at the future topology changes.

In the second half of the paper, we conduct an exhaustive simulation study of the Max.Stability-DG trees vs. the MST-DG trees and analyze their impact on the node lifetime, network lifetime and coverage loss time. To the best of our knowledge, we could not find any such comprehensive analysis of two data gathering strategies for mobile sensor networks and also with respect to the node failure times beyond the first node failure as well as analysis on the coverage loss time and fraction of coverage loss at any time due to node failures. The rest of the paper is organized as follows: Section 2 presents the algorithms to determine the Max.Stability-DG trees and MST-DG trees. Section 3 presents the simulation environment used and introduces the performance metrics. Section 4 describes the simulation results observed for the node and network lifetime. Section 5 describes the simulation results obtained for the coverage loss time and fraction of loss of coverage. Section 6 concludes the paper.

## 2. DATA GATHERING ALGORITHMS BASED ON MAXIMUM STABILITY AND MINIMUM DISTANCE SPANNING TREES

The system model adopted in this research is as follows: Each sensor node is assumed to operate with an identical and fixed *transmission range*. For the purpose of calculating the coverage loss, we also use the *sensing range* of a sensor node, considered in this research, as half the transmission range of the node. Basically, a sensor node can monitor and collect data at locations within the radius of its sensing range and transmit them to nodes within the radius of its transmission range. It has been proven in the literature [7] that the transmission range per node has to be at least twice the sensing range of the nodes to ensure that coverage implies connectivity. Data gathering proceeds in rounds. During a round of data gathering, data gets aggregated starting from the leaf nodes of the tree and propagates all the way to the leader node.

International Journal of Computer Networks & Communications (IJCNC) Vol.5, No.4, July 2013

An intermediate node in the tree collects the aggregated data from its immediate child nodes and further aggregates with its own data before forwarding to its immediate parent node in the tree.

We use the notions of static graphs and mobile graphs (adapted from [8]) to capture the sequence of topological changes in the network and determine a stable data gathering tree that spans over several time instants. A ***static graph*** is a snapshot of the network at any particular time instant and is modeled as a unit disk graph [9] wherein there exists a link between any two nodes if and only if the physical distance between the two end nodes of the link is less than or equal to the transmission range. The weight of an edge on a static graph is the Euclidean distance between the two end nodes of the edge. The Euclidean distance for a link $i - j$ between two nodes $i$ and $j$, currently at $(X_i, Y_i)$ and $(X_j, Y_j)$ is given by: $\sqrt{(X_i - X_j)^2 + (Y_i - Y_j)^2}$.

A ***mobile graph*** $G(i, j)$, where $1 \leq i \leq j \leq T$, where $T$ is the total number of rounds of the data gathering session corresponding to the network lifetime, is defined as $G_i \cap G_{i+1} \cap \ldots G_j$. Thus, a mobile graph is a logical graph that captures the presence or absence of edges in the individual static graphs. In this research work, we sample the network topology periodically for every round of data gathering to obtain the sequence of static graphs. The weight of an edge in the mobile graph $G(i, j)$ is the geometric mean of the weights of the edge in the individual static graphs spanning $G_i, \ldots, G_j$. Since there exist an edge in a mobile graph if and only if the edge exists in the corresponding individual static graphs, the geometric mean of these Euclidean distances would also be within the transmission range of the two end nodes for the entire duration spanned by the mobile graph. Note that at any time, a mobile graph includes only ***live sensor nodes***, nodes that have positive available energy.

## 2.1. Maximum Stability Spanning Tree-based Data Gathering (Max.Stability-DG) Algorithm

The Max.Stability-DG algorithm is based on a greedy look-ahead principle and the intersection strategy of static graphs. When a mobile data gathering tree is required at a sampling time instant $t_i$, the strategy is to find a mobile graph $G(i, j) = G_i \cap G_{i+1} \cap \ldots G_j$ such that there exists a spanning tree in $G(i, j)$ and no spanning tree exists in $G(i, j+1) = G_i \cap G_{i+1} \cap \ldots G_j \cap G_{j+1}$. We find such an epoch $t_i, \ldots, t_j$ as follows: Once a mobile graph $G(i, j)$ is constructed with the edges assigned the weights corresponding to the geometric mean of the weights in the constituent static graphs $G_i, G_{i+1}, \ldots, G_j$, we run the Prim's minimum-weight spanning tree algorithm on the mobile graph $G(i, j)$. If $G(i, j)$ is connected, we will be able to find a spanning tree in it. We repeat the above procedure until we reach a mobile graph $G(i, j+1)$ in which no spanning tree exists and there existed a spanning tree in $G(i, j)$. It implies that a spanning tree basically existed in each of the static graphs $G_i, G_{i+1}, \ldots, G_j$ and we refer to it as the mobile spanning tree for the time instants $t_i, \ldots, t_j$. To obtain the corresponding mobile data gathering tree, we choose an arbitrary root node for this mobile spanning tree and run the Breadth First Search (BFS) algorithm on it starting from the root node. The direction of the edges in the spanning tree and the parent-child relationships are set as we traverse its vertices using BFS. The resulting mobile data gathering tree with the chosen root node (as the leader node) is used for every round of data gathering spanning time instants $t_i, \ldots, t_j$. We then set $i = j+1$ and repeat the above procedure to find a mobile spanning tree and its corresponding mobile data gathering tree that exists for the maximum amount of time since $t_{j+1}$. A sequence of such maximum lifetime (i.e., longest-living) mobile data gathering trees over the timescale $T$ corresponding to the number of rounds of a data gathering session is referred to as the ***Stable Mobile Data Gathering Tree***. Figure 1 presents the pseudo code of the Max.Stability-DG algorithm that takes as input the sequence of static graphs spanning the entire duration of the data gathering session.

-----------------------------------------------------------------------------------------------------------------

**Input:** Sequence of static graphs $G_1, G_2, \ldots G_T$; Total number of rounds of the data gathering session – $T$
**Output:** *Stable-Mobile-DG-Tree*
**Auxiliary Variables:** *i, j*
**Initialization:** $i = 1$; $j=1$; *Stable-Mobile-DG-Tree* =

**Begin** *Max.Stability-DG Algorithm*

1  **while** ($i$   $T$) **do**

2     Find a mobile graph $G(i, j) = G_i \cap G_{i+1} \cap \ldots \cap G_j$ such that there exists at least one spanning tree in $G(i, j)$ and {no spanning tree exists in $G(i, j+1)$ or $j = T$}

3     *Mobile-Spanning-Tree*$(i, j)$ = **Prim's Algorithm** ( $G(i, j)$ )

4     *Root*$(i, j)$ = Choose a node randomly in $G(i, j)$

5     *Mobile-DG-Tree*$(i, j)$ = **Breadth First Search** ( *Mobile-Spanning-Tree*$(i, j)$, *Root*$(i, j)$ )

6     *Stable-Mobile-DG-Tree* = *Stable-Mobile-DG-Tree* U { *Mobile-DG-Tree*$(i, j)$ }

7     **for** each time instant $t_k \in \{t_i, t_{i+1}, \ldots, t_j\}$ **do**
      Use the *Mobile-DG-Tree*$(i, j)$ in $t_k$

8       **if** node failure occurs at $t_k$ **then**
        $j = k – 1$
        *break*
     **end if**
   **end for**

9     $i = j + 1$

10 **end while**

11 **return** *Stable-Mobile-DG-Tree*

**End** *Max.Stability-DG Algorithm*

-----------------------------------------------------------------------------------------------------------------
**Figure 1:** Pseudo Code for the Maximum Stability-based Data Gathering Tree Algorithm

While operating the algorithm under energy-constrained scenarios, one or more sensor nodes may die due to exhaustion of battery charge even though the underlying spanning tree may topologically exist. For example, if we have determined a data gathering tree spanning across time instants $t_i$ to $t_j$ using the above approach, and we come across a time instant $t_k$ ($i$   $k$   $j$) at which a node in the tree fails, we simply restart the Max.Stability-DG algorithm starting from time instant $t_k$ considering only the live sensor nodes (i.e., the sensor nodes that have positive available energy) and determine the longest-living data gathering tree that spans all the live sensor nodes since $t_k$. The pseudo code of the Max.Stability-DG algorithm in Figure 1 handles node failures, when run under energy-constrained scenarios, through the ***if*** block segment in statement 8. If all nodes have sufficient-energy and there are no node failures, the algorithm does not execute statement 8.

**2.2. Minimum Distance Spanning Tree based Data Gathering Algorithm**

In our simulation studies (sections 3 through 5), we compare the performance of the Max.Stability-DG trees with that of the minimum-distance spanning tree based data gathering (MST-DG) trees. The sequence of MST-DG trees for the duration of the data gathering session is generated as follows: If a MST-DG tree is not known for a particular round, we run the Prim's minimum-weight spanning tree algorithm on the static graph representing the snapshot of the network topology generated at the time instant corresponding to the round. Since the weights of the edges in a static graph represent the physical Euclidean distance between the constituent end nodes of the edges, the Prim's algorithm will return the minimum-distance spanning tree on the static graph. We then choose an arbitrary root node and run the Breadth First Search (BFS) algorithm starting from this node. The MST-DG tree is the rooted form of the minimum-distance spanning tree with the chosen root node as the leader node. We continue to use the MST-DG tree as long as it exists. The leader node of the MST-DG tree remains the same until the tree breaks due to node mobility or node failures. When the MST-DG tree ceases to exist for a round, we repeat the above procedure. This way, we generate a sequence of MST-DG trees, referred to as the ***MST Mobile Data Gathering Tree***. The MST-DG algorithm emulates the general strategy (referred to as Least Overhead Routing Approach, LORA [10]) of routing protocols and data gathering algorithms for ad hoc networks and sensor networks. That is, the algorithm chooses a data gathering tree that appears to be the best at the current time instant and continues to use it as long as it exists. In a recent work [5], we have observed the minimum-distance spanning tree-based data gathering trees to be the most energy-efficient communication topology for data gathering in static sensor networks. To be fair to the Max.Stability-DG algorithm that is proposed and evaluated in this research, the MST-DG algorithm is also run in a centralized fashion with the assumption that the entire static graph information is available at the beginning of each round.

## 3. SIMULATION ENVIRONMENT AND PERFORMANCE METRICS

We conduct an exhaustive simulation study on the performance of the Max.Stability-DG trees and compare them with that of the MST-DG trees under diverse conditions of network density and mobility. The simulations are conducted in a discrete-event simulator developed (in Java) by us exclusively for data gathering in mobile sensor networks. The MAC (medium access control) layer is assumed to be collision-free and considered an ideal channel without any interference. Sensor nodes are assumed to be both TDMA (Time Division Multiple Access) and CDMA (Code Division Multiple Access)-enabled [11]. Every upstream node broadcasts a time schedule (for data gathering) to its immediate downstream nodes; a downstream node transmits its data to the upstream node according to this schedule. Such a TDMA-based communication between every upstream node and its immediate downstream child nodes can occur in parallel, with each upstream node using a unique CDMA code.

The network dimension is 100m x 100m. The number of nodes in the network is 100 and initially, the nodes are uniform-randomly distributed throughout the network. The sink is located at (50, 300), outside the network field. For a given simulation run, the transmission range per sensor node is fixed and is the same across all nodes. The network density is varied by varying the transmission range per sensor node of 25m (representative of moderate density, with connectivity of 97% and above) and 40m (representative of high density, with 100% connectivity).

Each node is supplied with limited initial energy (2 J per node) and the simulations are conducted until the network of live sensor nodes gets disconnected due to the failures of one or more nodes. The energy consumption model used is a first order radio model [12] that has been also used in several of the well-known previous work (e.g., [1][2]) in the literature. According to this model, the energy expended by a radio to run the transmitter or receiver circuitry is $E_{elec}$ = 50 nJ/bit and $\in_{amp}$ = 100 pJ/bit/m$^2$ for the transmitter amplifier. The radios are turned off when a node wants to avoid receiving unintended transmissions. The energy lost in transmitting a *k*-bit message over a distance *d* is given by: $E_{TX}(k, d) = E_{elec}* k + \in_{amp}*k* d^2$. The energy lost to receive a *k*-bit message is: $E_{RX}(k) = E_{elec}* k$.

We conduct constant-bit rate data gathering at the rate of 4 rounds per second (one round for every 0.25 seconds). The size of the data packet is 2000 bits; the size of the control messages used for tree discoveries is assumed to be 400 bits. We assume that a tree discovery requires network-wide flooding of the 400-bit control messages such that each sensor node will broadcast the message exactly once in its neighborhood. As a result, each sensor node will lose energy to transmit the 400-bit message over its entire transmission range and receive the message from each of its neighbor nodes. In high density networks, the energy lost due to receipt of the redundant copies of the tree discovery control messages dominates the energy lost at a node for tree discovery. All of these mimic the energy loss observed for flooding-based tree discovery in ad hoc and sensor networks.

The node mobility model used is the well-known Random Waypoint mobility model [13] with the maximum node velocity being 3 m/s, 10 m/s and 20 m/s representing scenarios of low, moderate and high mobility respectively. According to this model, each node chooses a random target location to move with a velocity uniform-randomly chosen from [0,…, $v_{max}$], and after moving to the chosen destination location, the node continues to move by randomly choosing another new location and a new velocity. Each node continues to move like this, independent of the other nodes and also independent of its mobility history, until the end of the simulation. For a given $v_{max}$ value, we also vary the dynamicity of the network by conducting the simulations with a variable number of static nodes (out of the 100 nodes) in the network. The values for the number of static nodes used are: 0 (all nodes are mobile), 20, 50 and 80.
We generated 200 mobility profiles of the network for a total duration of 6000 seconds, for every combination of the maximum node velocity and the number of static nodes. Every data point in the results presented in Figures 2 through 11 is averaged over these 200 mobility profiles. The performance metrics measured in the simulations are:

(i) *Node Lifetime* – measured as the time of first node failure due to exhaustion of battery charge.

(ii) *Network Lifetime* – measured as the time of disconnection of the network of live sensor nodes (i.e., the sensor nodes that have positive available battery charge), while the network would have stayed connected if all the nodes were alive at that time instant. So, before confirming whether an observed time instant is the network lifetime (at which the network of live sensor nodes is noticed to be disconnected), we test for connectivity of the underlying network if all the sensor nodes were alive.
We obtain the distribution of node failures as follows: The probability for '*x*' number of node failures (*x* from ranging from 1 to 100 as we have a total of 100 nodes in our network for all the simulations) for a given combination of the operating conditions (transmission range per node, maximum node velocity and number of static nodes) is measured as the number of mobility profile files that reported *x* number of node failures divided by 200, which is the total number of mobility profiles used for every combination of maximum node velocity and number of static nodes. Similarly, we keep track of the time at which '*x*' (*x* ranging from 1 to 100) number of node failures occurred in each of the 200 mobility profiles for a given combination of operating conditions and the values for the time of node failures reported in Figures 4, 5 and 6 are an average of these data collected over all the mobility profile files. We discuss the results for the distribution of the time and probability of node failures along with the discussion on node lifetime and network lifetime in Section 4.

(iii) *Fraction of Coverage Loss and Coverage Loss Time*: If *f* is denoted as 'Fraction of Coverage Loss' (ranging from 0.01 to 1.0, measured in increments of 0.01), the coverage loss time is the time at which any *f* randomly chosen locations (X, Y co-ordinates) among 100 locations in the network is not within the sensing range of any node (explained in more detail below).

Since the number of node failures increases monotonically with time and network coverage depends on the number of live nodes, our assumption in the calculations for network coverage loss is that the fraction of coverage loss increases monotonically with time. We keep track of the largest fraction of coverage loss the network has incurred so far, and at the beginning of each round we check whether the network has incurred the next largest fraction of coverage loss, referred to as the *target fraction of coverage loss*. The first time instant during which we observe the network to have incurred the target coverage loss is recorded as the coverage loss time for the particular fraction of coverage loss, and from then on, we increment the target coverage loss by 0.01 and keep testing for the first occurrence of the new target fraction of coverage loss in the subsequent rounds. We repeat the above procedure until the network lifetime is encountered for the simulation with the individual data gathering algorithm.

At the beginning of each round, we check for network coverage as follows: We choose 100 random locations in the network and find out whether each of these locations is within the sensing range of at least one sensor node. We count the number of locations that are not within the sensing range of any node. If the fraction of the number of locations (actual number of locations that are not covered / total number of locations considered, which is 100) not within the sensing range of any node equals the target fraction of coverage loss, we record the time instant for that particular round of data gathering as the coverage loss time corresponding to the target fraction of coverage loss. We then increment the target fraction of coverage loss by 0.01 and repeat the above procedure to determine the coverage loss time corresponding to the new incremented value of the target fraction of coverage loss.
Each coverage loss time data point reported for particular fractions of coverage loss in Figures 9, 10 and 11 are the average values of the coverage loss times observed when the individual data gathering tree algorithms are run with the mobility profile files corresponding to a particular condition of network dynamicity (max. node velocity and number of static nodes) and transmission range per node. The probability for a particular fraction of coverage loss is computed as the ratio of the number of mobility profile files in which the corresponding fraction of coverage loss was observed divided by the total number of mobility profile files (200 mobility profile files for each operating condition).

## 4. NODE LIFETIME AND NETWORK LIFETIME

We observe a tradeoff between node lifetime and network lifetime for maximum stability vs. minimum-distance spanning tree based data gathering in mobile sensor networks. The MST-DG trees incur larger node lifetimes (the time of first node failure) for all the 48 operating combinations of maximum node velocity, number of static nodes and transmission range per node. The Max.Stability-DG trees incur larger network lifetime for most of the operating conditions. The lower node lifetime incurred with the Max.Stability-DG trees is attributed to the continued use of stable data gathering trees for a longer time and that too without changing the leader node. It would involve too much of message complexity and energy consumption to have the sensor nodes coordinate among themselves to choose a leader node for every round. Hence, we choose the leader node for a data gathering tree at the time of discovering it and let the leader node remain the same for the duration of the tree (i.e., until the tree fails). The same argument applies for the continued use of the intermediate nodes that receive aggregate data from one or more child nodes and transmit them to an upstream node in the tree. Due to the unfairness in node usage resulting from the overuse of certain nodes as intermediate nodes and leader node, the Max.Stability-DG trees have been observed to yield a lower node lifetime, especially under operating conditions (like low and moderate node mobility with moderate and larger transmission range per node) that facilitate greater stability.

The node lifetime incurred with the Max.Stability-DG trees increases significantly with increase in the maximum node velocity, especially when operated in moderate transmission ranges per node. We observe an increase in node lifetime by as large as 200-400% as we increase $v_{max}$ from 3 m/s to 10 m/s and operate the nodes at a moderate transmission range of 25m or 30m. A further increase in $v_{max}$ (i.e., from 10 m/s) to 20 m/s increases the node lifetime further by 50-100%. We did not observe an increase in node lifetime when we increase $v_{max}$ from 3 m/s to 10 m/s at higher transmission ranges per node of 40m. However, a further increase in the maximum node velocity to 20 m/s triggers regular tree failures that contribute to the fairness of node usage, resulting in an increase in node lifetime by 100-150%. A similar impact of node mobility on the node lifetime incurred with the MST-DG trees can also be observed, albeit at a lower percentage increase. We observe that the node lifetime for the MST-DG trees to increase by about 50-100% as we increase the maximum node velocity from 3 m/s to 10 m/s. However, a further increase in the maximum node velocity from 10 m/s to 20 m/s does not create a similar positive impact on the node lifetime; we observe the node lifetime to further increase by only about 10-20%, and in case of lower transmission ranges per node, we even observe a 5% decrease in node lifetime.

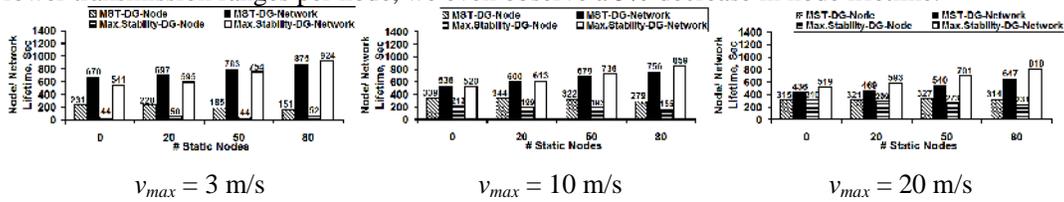

| $v_{max}$ = 3 m/s | $v_{max}$ = 10 m/s | $v_{max}$ = 20 m/s |

**Figure 2:** Average Node and Network Lifetime (Transmission Range per Node = 25 m)

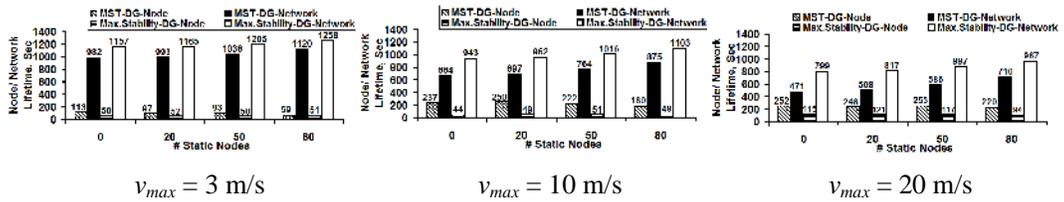

| $v_{max}$ = 3 m/s | $v_{max}$ = 10 m/s | $v_{max}$ = 20 m/s |

**Figure 3:** Average Node and Network Lifetime (Transmission Range per Node = 40 m)

The node lifetime incurred for the MST-DG trees can be larger than that of the Max.Stability-DG trees by as large as 400% at low and moderate levels of node mobility and by as large as 135% at higher levels of node mobility. For a given level of node mobility, the difference in the node lifetimes incurred for the MST-DG trees and Max.Stability-DG trees increases with increase in the transmission range per node (for a fixed number of static nodes) and either remain the same or slightly increase with increase in the number of static nodes (for a fixed transmission range per node). For a given level of node mobility, the node lifetime incurred with the Max.Stability-DG trees decreases by about 30-40% as we increase the transmission range per node from 25m to 30m, and decreases further by another 50-60% as we increase the transmission range per node from 30m to 40m. The MST-DG trees too suffer a decrease in node lifetime with increase in transmission range per node; but, at a lower scale – due to the relative instability of the trees. At larger transmission ranges per node, the data gathering trees are bound to be more stable, and the negative impact of this on node lifetime is significantly felt in the case of the Max.Stability-DG trees. For a given transmission range per node, the negative impact associated with the use of static nodes on node lifetime is increasingly observed at $v_{max}$ values of 3 m/s and 10 m/s. At $v_{max}$ = 20 m/s, since the network topology changes dynamically, even the use of 80 static nodes is not likely to overuse certain nodes and result in their premature failures. The node lifetime incurred with MST-DG trees is more impacted with the use of static nodes at low node mobility scenarios (Figure 4) and the node lifetime incurred with the Max.Stability-DG trees is more impacted with the use of static nodes at moderate and higher node mobility scenarios (Figures 5 and 6).

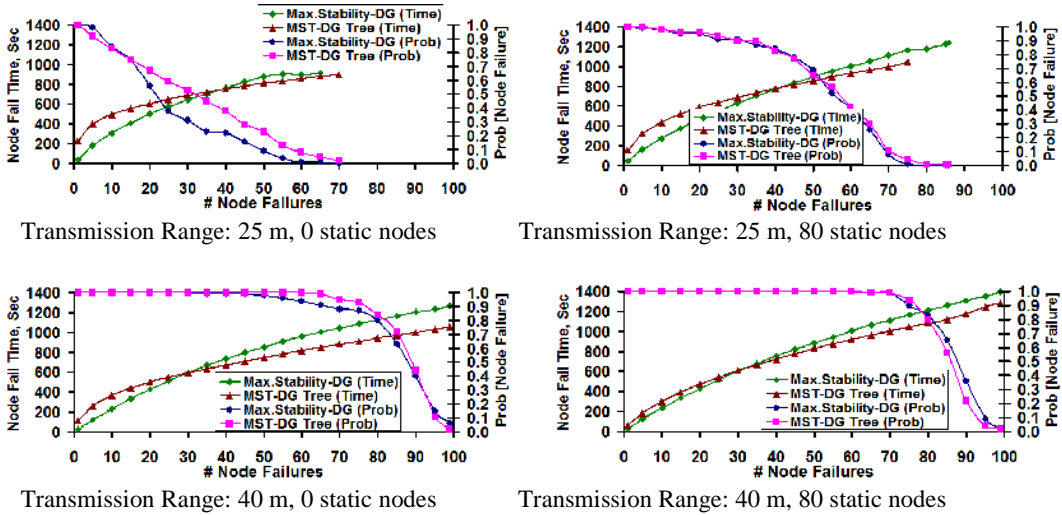

Transmission Range: 25 m, 0 static nodes   Transmission Range: 25 m, 80 static nodes

Transmission Range: 40 m, 0 static nodes   Transmission Range: 40 m, 80 static nodes

**Figure 4:** Distribution of Node Failure Times and Probability of Node Failures [$v_{max}$ = 3 m/s]

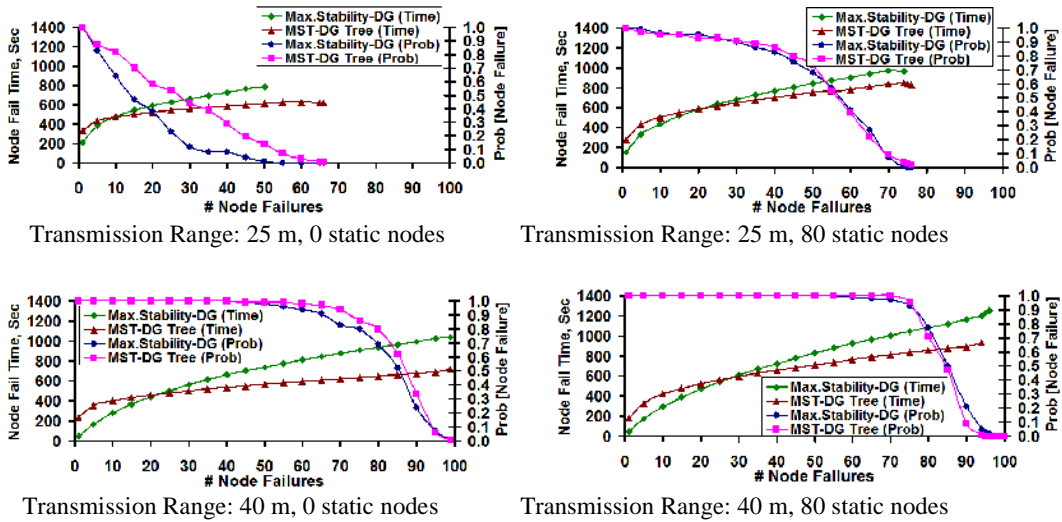

Transmission Range: 25 m, 0 static nodes   Transmission Range: 25 m, 80 static nodes

Transmission Range: 40 m, 0 static nodes   Transmission Range: 40 m, 80 static nodes

**Figure 5:** Distribution of Node Failure Times and Probability of Node Failures [$v_{max}$ = 10 m/s]

The Max.Stability-DG trees compensate for the premature failures of certain nodes by incurring a lower energy loss per round and energy loss per node due to lower tree discoveries and shorter tree height with more even distribution of the number of child nodes per intermediate node. As the dynamicity of the network increases, the data gathering trees become less stable, and this helps to rotate the roles of the intermediate nodes and leader node among the nodes to increase the fairness of node usage. All of these save significantly more energy at the remaining nodes that withstand the initial set of failures. As a result, we observe the Max.Stability-DG trees to observe a significantly longer network lifetime compared to that of the MST-DG trees. There are only four combinations of operating conditions under which the MST-DG trees incur larger network

lifetime – these correspond to transmission range per node of 25m and $v_{max}$ = 3 m/s (0, 20 and 50 static nodes), 20 m/s (0 static nodes).

The difference in the network lifetime incurred for the Max.Stability-DG trees and that of the MST-DG trees increases with increase in the maximum node velocity and transmission range per node. At low, moderate and high levels of node mobility, the network lifetime incurred with the Max.Stability-DG trees can be larger than that of the MST-DG trees by about 5-20%, 15-40% and 20-60% respectively, with the difference increasing with increase in the transmission range per node. Similar range of differences in the network lifetime can be observed for the two data gathering trees at transmission ranges per node of 25m, 30m and 40m, with the difference increasing as the maximum node velocity increases. For a given $v_{max}$ and transmission range per node, the number of static nodes does not make a significant impact on the difference in the network lifetime incurred with the two data gathering trees at moderate transmission ranges per node of 25 and 30m. However, at larger transmission ranges per node of 40m, the difference in the network lifetime decreases by about 15-35%. This could be attributed to the relatively high stability of the Max.Stability-DG trees when operated at larger transmission ranges per node in the presence of more static nodes.

With respect the impact of the operating parameters on the absolute magnitude of the network lifetime, we observe the network lifetime incurred with the two data gathering trees increases with increase in the number of static nodes for a given value of $v_{max}$ and transmission range per node. For a given level of node mobility, the network lifetime increases with increase in transmission range per node; however, for the MST-DG trees, the rate of increase decreases with increase in the maximum node velocity. This could be attributed to the relative instability of the MST-DG trees at high node mobility levels, requiring frequent tree reconfigurations. During a network-wide flooding, all nodes in the network tend to lose energy, almost equally. The Max.Stability-DG trees maintain a steady increase in the network lifetime with increase in transmission range per node for all levels of node mobility. For a given transmission range per node and number of static nodes, the network lifetime incurred for the two data gathering trees decreases with increase in the maximum node velocity, especially for the MST-DG trees due to their instability. This could be attributed to the energy loss incurred due to frequent tree discoveries. The network lifetime incurred with the Max.Stability-DG trees and MST-DG trees decreases by about 30-50% and 50-100% respectively as we increase the maximum node velocity from 3 m/s to 20 m/s for a fixed transmission range per node and number of static nodes.

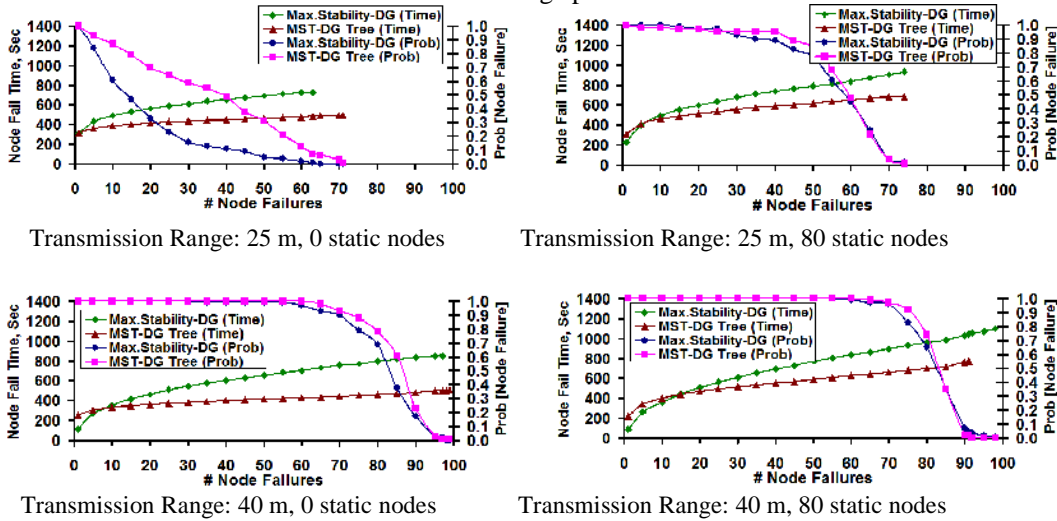

| Transmission Range: 25 m, 0 static nodes | Transmission Range: 25 m, 80 static nodes |
| Transmission Range: 40 m, 0 static nodes | Transmission Range: 40 m, 80 static nodes |

**Figure 6:** Distribution of Node Failure Times and Probability of Node Failures [$v_{max}$ = 20 m/s]

For a given transmission range per node, with the absolute values of the node lifetime increasing with increase in the maximum node velocity and the network lifetime decreasing with increase in the maximum node velocity, we observe the maximum increase in the absolute time of node failures to occur at low node mobility. This vindicates the impact of network-wide flooding based tree discoveries on energy consumption at the nodes. Since all nodes are likely to lose the same amount of energy with flooding, the more we conduct flooding, the larger is the network-wide energy consumption. As a result, node failures tend to occur more frequently when we conduct frequent flooding. Thus, even though operating the network at moderate and high levels of node mobility helps us to extend the time of first node failure, the subsequent node failures occur too soon after the first node failure. This could be justified with the observation of flat curves for the MST-DG trees with respect to the distribution of node failure times (in Figures 4, 5 and 6). The distribution of node failure times is relatively steeper for the Max.Stability-DG trees. The unfair usage of nodes in the initial stages does help the Max.Stability-DG trees to prolong the network lifetime. Aided with node mobility, it is possible for certain energy-rich nodes (that might have been leaf nodes in an earlier data gathering tree) to keep the network connected for a longer time by serving as intermediate nodes, and the energy-deficient nodes serve as leaf nodes during the later rounds of data gathering.

The impact of mobility in prolonging node failure lifetimes could also be explained by the lower probability of node failure observed for the Max.Stability-DG trees in comparison to the MST-DG trees when there are 0 static nodes (the plots to the left in Figures 4, 5 and 6). At 80 static nodes, the probability of node failures for the two data gathering trees is about the same and is higher than that observed when all nodes are mobile. This could be attributed to the repeated overuse of certain nodes as intermediate nodes and leader node on relatively more stable data gathering trees. Thus, with the use of static nodes, even though the absolute magnitude of the network lifetime can be marginally increased (by about 10-70%; the increase is larger at moderate transmission range per node and larger values of $v_{max}$), the probability of node failures to occur also increases.

In terms of the percentage difference in the values for the network lifetime and node lifetime incurred with the two data gathering trees, we observe the Max.Stability-DG trees to incur a significantly prolonged network lifetime, beyond the time of first node failure. For a given transmission range per node and maximum node velocity, we observe the difference between the node lifetime and network lifetime for the Max.Stability-DG trees to increase significantly with increase in the number of static nodes. This could be attributed to the reduction in the number of flooding-based tree discoveries. For a given level of node mobility, we observe the difference in the node lifetime and network lifetime for the Max.Stability-DG trees to increase with increase in the transmission range per node. This could be again attributed to the decrease in the number of network-wide flooding based tree discoveries when operated at larger transmission ranges per node. Relatively, the MST-DG trees incur a very minimal increase in the network lifetime compared to the node lifetime, especially when operated at higher levels of node mobility. The network lifetime incurred with the Max.Stability-DG trees could be larger than the node lifetime as low as by a factor of 1.7 and as large as by a factor of 23. On the other hand, the network lifetime incurred with the MST-DG trees could be larger than the node lifetime as low as by a factor of 1.4 and as large as by a factor of 5.7.

One can also observe from Figures 4, 5 and 6 that the number of node failures that require for the node failure time incurred with the Max.Stability-DG trees to exceed that of the node failure time incurred with the MST-DG trees decreases with increase in maximum node mobility. This could be attributed to the premature very early node failure occurring for the Max.Stability-DG trees when operated under low node mobility scenarios, with the time of first node failure for the MST-DG tree being as large as 400% more than the time of first node failure for the Max.Stability-DG tree. On the other hand, at high levels of node mobility, the time of first node failure incurred with the MST-DG trees is at most 100% larger than that of the Max.Stability-DG trees. Hence, the

International Journal of Computer Networks & Communications (IJCNC) Vol.5, No.4, July 2013

node failure times incurred with the Max.Stability-DG trees could quickly exceed that of the MST-DG trees at higher levels of node mobility. At the same time, the probability for node failures to occur (that was relatively low at moderate transmission ranges per node, low and moderate levels of node mobility) with the Max.Stability-DG trees converges to that of the MST-DG trees when operated at higher levels of node mobility as well as with larger transmission ranges per node. For a given $v_{max}$ value and transmission range per node, we also observe that the number of node failures required for the failure times incurred with the Max.Stability-DG trees to exceed that of the MST-DG trees increases with increase in the number of static nodes.

## 5. COVERAGE LOSS ANALYSIS
### 5.1 Coverage Loss at a Common Timeline

In this section, we compare the loss of coverage incurred with both the Max.Stability-DG and MST-DG trees with respect to a ***common timeline, chosen to be the minimum of the network lifetime obtained for the two data gathering trees*** under every operating condition of transmission range per node, maximum node velocity and the number of static nodes. Given the nature of the results obtained for the network lifetime under different operating conditions, the minimum of the network lifetime for the two data gathering trees ended up mostly being the network lifetime observed for the MST-DG trees. For this value of network lifetime, we measured the fraction of coverage loss in the network incurred for each of the two data gathering trees, as well as measured the probability with which the corresponding fraction of coverage loss is observed.

Under the above measurement model, we observe the Max.Stability-DG trees incur lower values of the fractions of coverage loss at the minimum of the network lifetime incurred for the two data gathering trees for most of all the 48 combinations of the operating conditions of maximum node velocity, number of static nodes and transmission range per node (see Figures 7 and 8). However, the fraction of coverage loss observed for the Max.Stability-DG trees is bound to occur with a higher probability than that of the coverage loss to be incurred by using the MST-DG trees. The difference in the fraction of coverage loss incurred for the Max.Stability-DG trees vis-à-vis could be as large as 0.18-0.21, observed at transmission range per node of 40m and 80 static nodes, under all levels of node mobility. The only three combinations of operating conditions for which the Max.Stability-DG trees sustain a larger value for the fraction of coverage loss (that too, only by 0.02) are at a transmission range per node of 25m - $v_{max}$ = 3 m/s, 0 and 20 static nodes; and $v_{max}$ = 10 m/s, 0 static nodes.

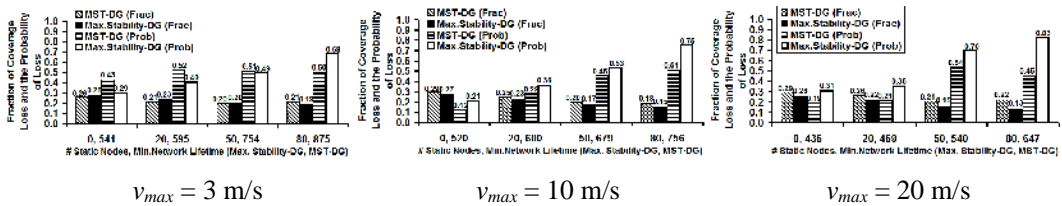

$v_{max}$ = 3 m/s    $v_{max}$ = 10 m/s    $v_{max}$ = 20 m/s
**Figure 7:** Fraction of Coverage Loss and Associated Probability (Trans. Range/Node = 25 m)

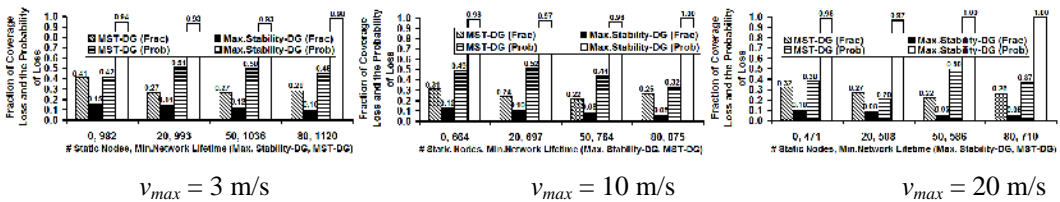

$v_{max}$ = 3 m/s    $v_{max}$ = 10 m/s    $v_{max}$ = 20 m/s
**Figure 8:** Fraction of Coverage Loss and Associated Probability (Trans. Range/Node = 40 m)

In the case of the Max.Stability-DG trees, for a fixed $v_{max}$ value, we observe the fraction of loss of coverage to decrease with increase in transmission range per node from 25m to 40m, of course with a higher probability. The significant decrease in the loss of coverage (as low as 0.05) at higher transmission range per node of 40m could also be attributed to the increase in the network lifetime, and also due to the reason that we measure the loss of coverage at a time value (corresponding to the network lifetime of the MST-DG trees), which is lower than the network lifetime of the Max.Stability-DG trees. For fixed $v_{max}$ and transmission range, as we increase the number of static nodes, the fraction of coverage loss decreases significantly for Max.Stability-DG trees by about 0.05 to 0.1; whereas, the fraction of coverage loss for the MST-DG trees suffers a very minimal decrease or remains the same. For a fixed # static nodes and transmission range per node, the maximum node velocity has minimal impact on the fraction of coverage loss for the MST-DG trees. However, for the Max.Stability-DG trees, as we increase the maximum node velocity from 3 m/s to 20 m/s, at transmission ranges per node of 30m and 40m, the fraction of coverage loss that was already low (in the range 0.10 – 0.15 range) decreases further by about 0.05 – 0.07.

### 5.2 Distribution of Coverage Loss

In Figures 9, 10 and 11, we illustrate the distribution of the time (referred to as the coverage loss time) at which particular fractions of coverage loss occurs in the network when run with the Max.Stability-DG and MST-DG trees (until the network lifetime of the individual data gathering tree). The Max.Stability-DG trees incur larger values of coverage loss time for moderate and higher values of the fractions of coverage loss (generally above 0.15 or 0.2), under most of the combinations of the operating conditions of maximum node velocity, 0 and 80 static nodes and transmission range per node. The only combination of operating conditions for which the Max.Stability-DG trees sustain a lower coverage loss time for fractions of coverage loss greater than 0.15-0.2 is at $v_{max} = 3$ m/s – transmission ranges per node of 25m. For quantitative comparison purposes, we base our discussion in this section on the coverage loss time observed when the fraction of coverage loss is 0.3. For most of the combinations of operating conditions, we observe the coverage loss times incurred with the Max.Stability-DG and MST-DG trees to flatten out (i.e., not appreciably increase) starting from this fraction of coverage loss.

In terms of the percentage difference in the coverage loss time incurred at a fraction of coverage loss of 0.3, we observe the coverage loss time incurred with the Max.Stability-DG trees to be about 15-40%, 15-45% and 30-70% greater than the coverage loss time incurred with the MST-DG trees at low, moderate and high levels of node mobility respectively. For a fixed transmission range per node and number of static nodes, the absolute magnitude for the coverage loss time incurred for both the data gathering trees decreases with increase in the $v_{max}$ value. The MST-DG trees suffer the most with their coverage loss time decreasing by about 20-40% as we increase $v_{max}$ from 3 m/s to 10 m/s, and a further decrease by another 20-40% as we increase $v_{max}$ from 10 m/s to 20 m/s. The Max.Stability-DG trees suffer a relatively slower decrease in coverage loss time by about 10-25% as we increase $v_{max}$ from 3 m/s to 10 m/s, and a further decrease by another 5-15% as we increase $v_{max}$ from 10 m/s to 20 m/s. The consolation is that the decrease in the coverage loss time occurs at a lower probability for the fraction of coverage loss to be at 0.3.

To illustrate the immense negative impact of increase in the maximum node velocity on the coverage loss time for MST-DG trees, we cite the following observation from Figures 9, 10 and 11: the coverage loss time incurred with $v_{max} = 3$ m/s, transmission range of 25m and 0 static nodes is even greater than the coverage loss time incurred with $v_{max} = 20$ m/s, transmission range of 40m and 80 static nodes (by about 5%). Thus, neither the increase in the number of static nodes and nor the increase in the transmission range per node can adequately compensate for the decrease in the coverage loss time when we increase the maximum velocity of a node from 3 m/s to 20 m/s. On the contrary, for the same conditions, we observe the coverage loss time incurred

International Journal of Computer Networks & Communications (IJCNC) Vol.5, No.4, July 2013

with the Max.Stability-DG trees at $v_{max}$ = 20 m/s, transmission range per node of 40m and 80 static nodes to be about 70% greater than the coverage loss time incurred at $v_{max}$ = 3m/s, transmission range of 25m and 0 static nodes. This emphasizes the importance of discovering stable data gathering trees that can reduce the energy lost due to frequent network-wide flooding at high levels of node mobility and prolong the coverage loss time incurred with the data gathering trees.

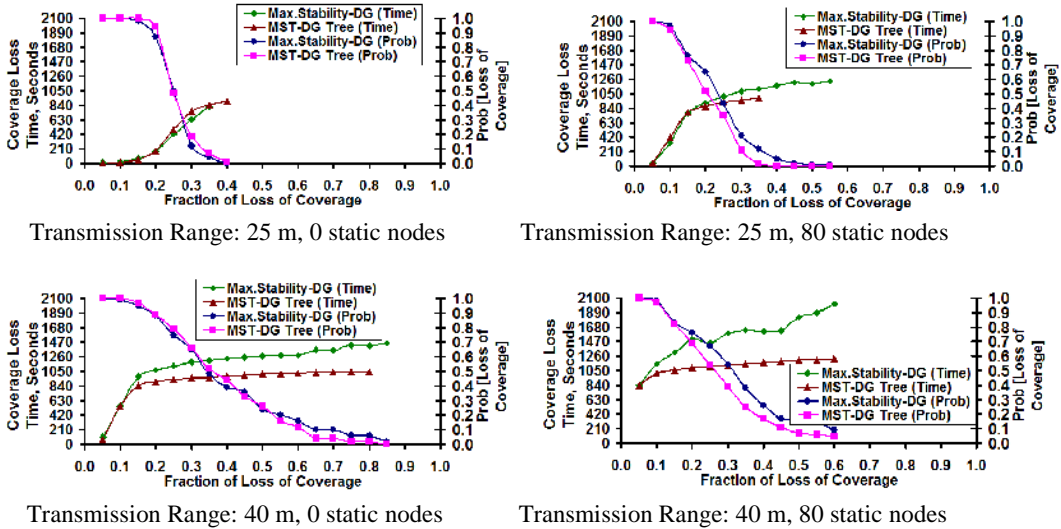

Transmission Range: 25 m, 0 static nodes  Transmission Range: 25 m, 80 static nodes

Transmission Range: 40 m, 0 static nodes  Transmission Range: 40 m, 80 static nodes

**Figure 9:** Coverage Loss Time and Probability of Coverage Loss [Low Mobility: $v_{max}$ = 3 m/s]

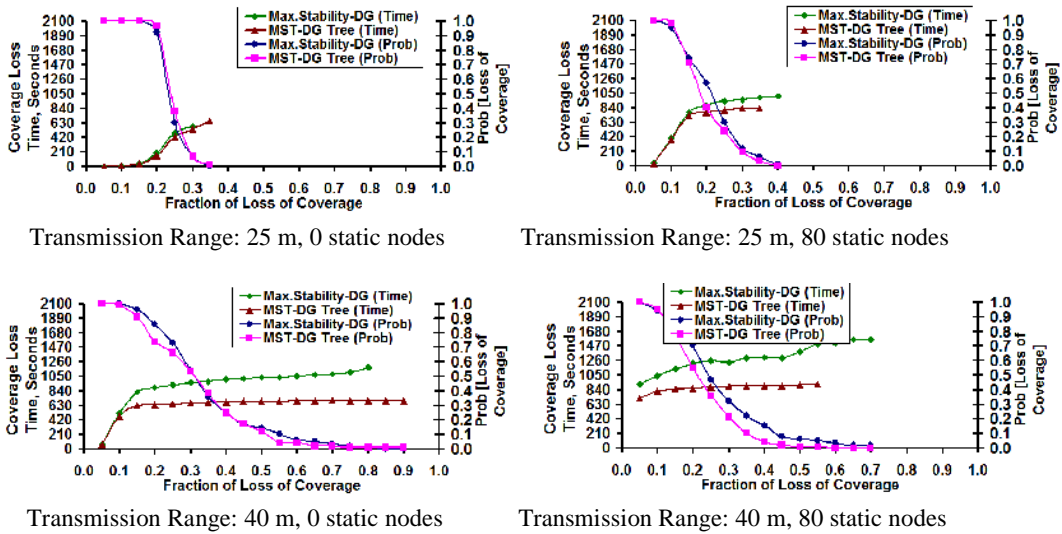

Transmission Range: 25 m, 0 static nodes  Transmission Range: 25 m, 80 static nodes

Transmission Range: 40 m, 0 static nodes  Transmission Range: 40 m, 80 static nodes

**Figure 10:** Coverage Loss Time & Probability of Coverage Loss [Mod. Mobility: $v_{max}$ = 10 m/s]

For a given level of node mobility, the coverage loss time incurred with the Max.Stability-DG trees almost doubles, if not more, as we increase the transmission range per node from 25m to 40m and the number of static nodes from 0 to 80. This could be attributed to the significant energy savings obtained as a result of the need for very few network-wide flooding tree discoveries with the use of the Max.Stability-DG algorithm when operated at larger transmission ranges per node and/or more static nodes. We observe significant gains in the coverage loss time when the number of static nodes is also simultaneously increased with increase in the transmission range per node. In fact, at moderate and high levels of node mobility, the coverage loss time incurred when we run the network at transmission range per node of 25m and increase the number of static nodes from 0 to 80 is greater than (by about 10%) or equal to the coverage loss time incurred when we run the network with 0 static nodes and increase the transmission range per node from 25m to 40m. In the case of MST-DG trees, the percentage increase in the coverage loss time with increase in the number of static nodes vis-à-vis increase in the transmission range per node is more obvious. The coverage loss time observed while using the MST-DG tree at a transmission range per node of 25m and increasing the number of static nodes from 0 to 80 is about 25-45% greater than the coverage loss time incurred when we run the network with 0 static nodes and increase the transmission range per node from 25m to 40m.

For both the data gathering trees, especially in the case of MST-DG trees, the potential energy savings obtained with respect to reduction in the number of network-wide flooding discoveries is much more when we operate at a moderate transmission range per node and increase the number of static nodes from 0 to 80 rather than operating at a larger transmission range per node with 0 static nodes. It is to be noted that larger the transmission range, the larger is the energy lost in transmission, and also larger is the energy lost due to receipt of the control messages from a larger number of neighbor nodes. For both the data gathering trees, we observe the increase in coverage loss time with the use of more static nodes vis-à-vis a larger transmission range per node to occur with a relatively lower probability of coverage loss.

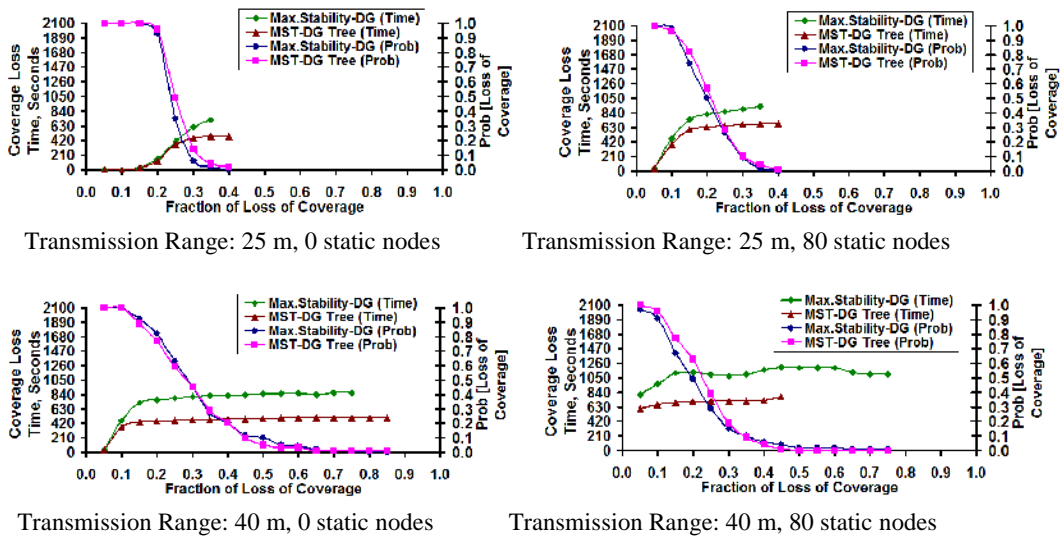

Transmission Range: 25 m, 0 static nodes    Transmission Range: 25 m, 80 static nodes

Transmission Range: 40 m, 0 static nodes    Transmission Range: 40 m, 80 static nodes

**Figure 11:** Coverage Loss Time & Probability of Coverage Loss [High Mobility: $v_{max} = 20$ m/s]

## 6. CONCLUSIONS

We have conducted an extensively evaluated the performance of the Max.Stability-DG trees and MST-DG trees under diverse conditions of network dynamicity (varied by changing the maximum node velocity and number of static nodes) and network density (varied by changing the transmission range per node). Due to its nature to use a long-living data gathering tree as long as it exists, we observe the Max.Stability-DG algorithm to incur a lower time for the first node failure. However, the tradeoff between stability and fairness of node usage ceases to exist beyond the first few node failures; the reduced number of network-wide flooding discoveries coupled with the shallow structure and even distribution of nodes across the intermediate nodes (which also contribute to a lower delay per round) contribute to a longer lifetime for the remaining nodes in the network and significantly prolong the network lifetime as well as the coverage loss time. On the contrary, the MST-DG trees that incur a larger time for the first node failure are observed to incur a significantly lower network lifetime and lower coverage loss time for a given fraction of loss of coverage (and correspondingly incur a larger fraction of coverage loss at any time), owing to frequent network-wide flooding-based tree discoveries that expedite the node failures after the first node failure. We did not come across such a comprehensive analysis for node failure times, network lifetime, coverage loss times and fraction of coverage loss in any prior work in the literature.

## ACKNOWLEDGMENTS

This research was sponsored by the U. S. Air Force Office of Scientific Research (AFOSR) through the Summer Faculty Fellowship Program for the lead author (Natarajan Meghanathan) in June-July 2012. The research was conducted under the supervision of the co-author (Philip D. Mumford) at the U. S. Air Force Research Lab (AFRL), Wright-Patterson Air Force Base (WPAFB) Dayton, OH. The AFRL public release number for this article is 88ABW-2012-4893. The views and conclusions in this document are those of the authors and should not be interpreted as representing the official policies, either expressed or implied, of the funding agency.

## REFERENCES

[1] S. Lindsey, C. Raghavendra and K. M. Sivalingam, "Data Gathering Algorithms in Sensor Networks using Energy Metrics," IEEE Transactions on Parallel and Distributed Systems, 13(9):924-935, September 2002.
[2] W. Heinzelman, A. Chandrakasan and H. Balakarishnan, "Energy-Efficient Communication Protocols for Wireless Microsensor Networks," Proceedings of the Hawaiian International Conference on Systems Science, January 2000.
[3] N. Meghanathan, "An Algorithm to Determine Energy-aware Maximal Leaf Nodes Data Gathering Tree for Wireless Sensor Networks," Journal of Theoretical and Applied Information Technology, , vol. 15, no. 2, pp. 96-107, May 2010.
[4] N. Meghanathan, "A Data Gathering Algorithm based on Energy-aware Connected Dominating Sets to Minimize Energy Consumption and Maximize Node Lifetime in Wireless Sensor Networks," International Journal of Interdisciplinary Telecommunications and Networking, vol. 2, no. 3, pp. 1-17, July-September 2010.
[5] N. Meghanathan, "A Comprehensive Review and Performance Analysis of Data Gathering Algorithms for Wireless Sensor Networks," International Journal of Interdisciplinary Telecommunications and Networking, vol. 4, no. 2, pp. 1-29, April-June 2012.
[6] T. H. Cormen, C. E. Leiserson, R. L. Rivest and C. Stein, "Introduction to Algorithms," 3rd Edition, MIT Press, July 2009.
[7] H. Zhang and J. C. Hou, "Maintaining Sensing Coverage and Connectivity in Large Sensor Networks," Wireless Ad hoc and Sensor Networks: An International Journal, vol. 1, no. 1-2, pp. 89-123, January 2005.
[8] A. Farago and V. R. Syrotiuk, "MERIT: A Scalable Approach for Protocol Assessment," Mobile Networks and Applications, Vol. 8, No. 5, pp. 567 – 577, October 2003.
[9] F. Kuhn, T. Moscibroda and R. Wattenhofer, "Unit Disk Graph Approximation," Proceedings of the Workshop on Foundations of Mobile Computing, pp. 17-23, October 2004.
[10] M. Abolhasan, T. Wysocki and E. Dutkiewicz, "A Review of Routing Protocols for Mobile Ad hoc Networks," Ad hoc Networks, vol. 2, no. 1, pp. 1-22, January 2004.
[11] A. J. Viterbi, "CDMA: Principles of Spread Spectrum Communication," 1st edition, Prentice Hall, April 1995.
[12] T. S. Rappaport, "Wireless Communications: Principles and Practice," 2nd edition, Prentice Hall, January 2002.
[13] C. Bettstetter, H. Hartenstein and X. Perez-Costa, "Stochastic Properties of the Random-Way Point Mobility Model," Wireless Networks, vol. 10, no. 5, pp. 555 – 567, September 2004.